\documentclass[12pt]{article}
\makeatletter
\@addtoreset{equation}{section}
\makeatother

\newcommand{\be}{\begin{equation}}
\newcommand{\ee}{\end{equation}}
\newcommand{\bea}{\begin{eqnarray}}
\newcommand{\eea}{\end{eqnarray}}
\def\dd{\partial}

\begin{document}

\begin{titlepage}

\begin{center}
\vskip 2.5 cm
{\Large \bf New duality transformation in two-dimensional non-linear sigma models}
\vskip 1 cm
{\large 
N. Mohammedi\footnote{e-mail:\ \tt nouri@celfi.phys.univ-tours.fr}}\\
\vskip 1cm
{\em Laboratoire de Math\'ematiques et Physique Th\'eorique\footnote{CNRS UMR 6083}\\
Universit\'e {}Fran\c{c}ois Rabelais \\
{}Facult\'e des Sciences et Techniques \\
Parc de Grandmont \\
{}F-37200 Tours - {}France.} 
\end{center}
\vskip 1.5 cm

\begin{abstract} 
\noindent
A new T-duality transformation is found in two-dimensional non-linear sigma models. 
Abelian and non-Abelian T-dualities are special cases of this construction.

\end{abstract}

\end{titlepage}

       \section{Introduction}

The duality transformations, known as T-duality, have proven to be a
crucial step in the study of string theory. These transformations 
stipulate that string propagation in some background could be 
equally described by another, a priori, completely different background.
Therefore, T-duality transformations are a useful tool in the programme
of classifying equivalent compactifications of string theory.
These transformations are constructed at the level of the two-dimensional
non-linear sigma model underlying the propagation of strings in non-trivial
backgrounds. There are, however, no systematic criteria for finding these 
duality transformations.
\par
The first T-duality transformations (abelian T-duality) were constructed for non-linear sigma 
models possessing Abelian isometries \cite{buscher}. This study was later generalised 
to theories with non-Abelian isometries (non-Abelian T-duality) \cite{quevedo,aagl}. It turned out, however, that 
the presence of isometries is not an essential ingredient in building T-duality transformations.
This is certainly what happens in the case of Poisson-Lie T-duality \cite{klimcik}.
These last transformations do not rely on isometries and are indeed symmetries of the 
string effective action \cite{our2}. 
\par
The idea behind Poisson-Lie T-duality resides in the fact that if two theories 
are dual then the equations of motion in one theory become Bianchi identities 
in the other (see \cite{poisson-lie} for various developments of this subject). 
In this letter, we exploite this observation to construct new T-duality transformations.
These are obvious generalisations of Abelian and non-Abelian T-dualities.  
  
\par
In order to clarify our approach, let us start by briefly revisiting 
Abelian and non-Abelian T-dualities. 
We consider the two-dimensional non-linear sigma model as represented by the 
action{\footnote{The world-sheet is for simplicity taken to be flat and we are 
using complex coordinates $z$ and $\bar{z}$.}}
\begin{eqnarray}
S\left[x,y\right] &=&  \int {\rm{d}}z{\rm{d}}\bar{z}\left\{
 Q_{ij}\left(x\right)\partial x^i \bar\partial x^j
+G_{ia}\left(x\right)\partial x^i \bar{A}^a  
+\bar G_{ia}\left(x\right)\bar\partial x^i A^a \right. \nonumber \\
 && +  
\left. P_{ab}\left(x\right) A^a \bar{A}^b \right\}\,\,\,.
\label{original} 
\end{eqnarray}
The target space coordinates are the ``spectators'' $x^i$ (labelled by $i,j,k,...$) and the
``active'' coordinates $y^a$ (labelled by $a,b,c,d,e,f,g,...$). The most important feature of these 
sigma models is that the $y^a$ coordinates appear only through
the one form 
\begin{equation}
\left\{
\begin{array}{c}
A^a= e^a_b\left(y\right)\partial y^b\\
\bar{A}^a= e^a_b\left(y\right)\bar\partial y^b
\end{array}
\right.
\,\,\,\,\,\,\,\,\,\,\,{\rm{or}}\,\,\,\,\,\,\,\,\,\,\,
\left\{
\begin{array}{c}
\partial y^a= E^a_b\left(y\right) A^b\\
\bar\partial y^a= E^a_b\left(y\right)\bar A^b 
\end{array}
\right.
\label{dy}
\end{equation}
Here $e^a_i$ are vielbeins satisfying the Cartan-Maurer relation 
$\partial _{a}e_{b}^{c}-\partial _{b}e_{a}^{c} + f_{dg}^{c}e_{a}^{d}e_{b}^{g}=0$,
where $f^a_{bc}$ are the structure constants of a Lie algebra $\cal{G}$. 
The inverses of these vielbeins are denoted $E^a_b$ and satisfy the commutation
relations $E^c_a\dd_c E^d_b -E^c_b\dd_c E^d_a -f^c_{ab}E^d_c=0$.
\par
The next ingredient in the construction of the dual theory is the integrability condition
(the Bianchi identity) associated 
with the definition (\ref{dy}). Namely;   
$\partial\bar\partial y^a- \bar\partial\partial y^a=0$. When expressed in terms of
$A^a$ and $\bar A^a$ it reads \cite{amit} 
\begin{equation}
E^a_b \left(\partial \bar A^{b} - \bar \partial A^{b}
+f_{cd}^{b}A^{c}\bar A^{d}\right)=0\,\,\,.
\label{bianchi}
\end{equation}
The crucial point about this Bianchi identity is that there is no explicit dependence 
on the coordinates $y^a$ in the expression between brackets.
\par
The Bianchi identity allows one to formulate the non-linear sigma model
(\ref{original}) in a first order formalism through the action \cite{townsend}
\begin{eqnarray}
S_1\left[x,A,\bar A, \chi\right] & = & 
\int {\rm{d}}z{\rm{d}}\bar{z}\left\{
Q_{ij}\left(x\right)\partial x^i \bar\partial x^j
+G_{ia}\left(x\right)\partial x^i \bar{A}^a  
+\bar G_{ia}\left(x\right)\bar\partial x^i A^a \right. \nonumber \\
&& + 
\left. P_{ab}\left(x\right) A^a \bar{A}^b 
+ \chi_b\left(\partial \bar A^{b} - \bar \partial A^{b}
+f_{cd}^{b}A^{c}\bar A^{d}\right)
\right\}\,\,\,.
\label{firstorder} 
\end{eqnarray}
This last action is equivalent to (\ref{original}) as the equations of motion 
for the Lagrange multiplier $\chi_a$ is simply the Bianchi identity (\ref{bianchi})
whose solution is the one form (\ref{dy}). Replacing this solution in the 
first order action leads to our starting theory (\ref{original}).
The dual theory is obtained by keeping the Lagrange multiplier $\chi_a$
and eliminating, instead, the independent fields 
$A^a$ and $\bar A^a$ through their equations of motion. The resulting two-dimensional
non-linear sigma model (the dual model) is described by the action 
\begin{eqnarray}
\widetilde{S}\left[x,\chi\right] &=&  \int {\rm{d}}z{\rm{d}}\bar{z}\left\{
\left[Q_{ij} - \left(M^{-1}\right)^{ab} G_{ia} \bar G_{jb}\right]
\partial x^i \bar\partial x^j
- \left(M^{-1}\right)^{ab} G_{ia} \dd x^i \bar\dd\chi_b
\right.\nonumber\\
&&+\left(M^{-1}\right)^{ba} \bar{G}_{ia} \bar\dd x^i \dd\chi_b
+\left(M^{-1}\right)^{ab}\partial\chi_a\bar\partial\chi_b
\left.\right\}\,\,\,,
\label{dual} 
\end{eqnarray} 
where the matrix $M_{ab}$ is defined as
\begin{equation}
M_{ab}= P_{ab} +\chi_c f^c_{ab}\,\,\,.
\end{equation}
In the dual theory the field $\chi_a$ plays the role of the field $y^a$.
It is worth mentioning that the two non-linear sigma models lead 
to the same string effective action as shown in \cite{our1}.
This construction will be generalised below in an obvious manner.

\section{The construction}

The important point in the above construction is the existence of the Bianchi
identity related to the definition (\ref{dy}). One notices, however,  that
$A^a$ and $\bar A^a$ are functions of the coordinates $y^a$ only. The idea behind
the new duality is to allow $A^a$ and $\bar A^a$ to depend on both $y^a$ and
$x^i$. We therefore take the new one-form to be defined by
\begin{equation}
\left\{
\begin{array}{c}
\partial y^a= \alpha^a_b\left(x,y\right)A^b+\beta^a_i\left(x,y\right)\partial x^i\\
\bar \partial y^a= \gamma^a_b\left(x,y\right)\bar A^b+\omega^a_i\left(x,y\right)\bar\partial x^i
\end{array}
\right.
\,\,\,\,\,\,\,\,\,\,\,{\rm{or}}\,\,\,\,\,\,\,\,\,\,\,
\left\{
\begin{array}{c}
A^a= \left(\alpha^{-1}\right)^a_b\left[\partial y^b-\beta^b_i\partial x^i\right]\\
\bar A^a= \left(\gamma^{-1}\right)^a_b\left[\bar\partial y^b-\omega^b_i\bar\partial x^i\right]
\end{array}
\right.
\label{newdy}
\end{equation} 
In terms of this one-form, the Bianchi identity corresponding to 
$\partial\bar\partial y^a- \bar\partial\partial y^a=0$ takes the form 
\begin{equation}
\gamma^a_b\left(
\partial\bar A^b - \theta^b_c\bar\partial A^c
+ \varphi^b_{cd}A^c\bar A^d 
+ \mu^b_{ic}\bar\partial x^i A^c 
+\nu^b_{ic}\partial x^i \bar A^c
+ \xi^b_{ij}\partial x^i \bar\partial x^j
+\rho^b_i\partial \bar \partial x^i\right)= 0 \,\,\,.
\label{newbianchi}
\end{equation}
The different tensors appearing in this equation are defined as 
\begin{eqnarray}
\theta^a_b &=& \left(\gamma^{-1}\right)^a_e\alpha^e_b
\label{one} \,\,\,\,\,,\\
\varphi^a_{bc} &=& 
\left(\gamma^{-1}\right)^a_e 
\left[\alpha^d_b\partial_d\gamma^e_c-\gamma^d_c\partial_d\alpha^e_b\right]
\label{two} \,\,\,\,\,,\\
\mu^a_{ib} &=& -\left(\gamma^{-1}\right)^a_e 
\left[\partial_i \alpha^e_b +\omega^c_i\partial_c\alpha^e_b 
- \alpha^c_b\partial_c \omega^e_i\right] 
\label{three} \,\,\,\,\,,\\
\nu^a_{ib} &=& \left(\gamma^{-1}\right)^a_e 
\left[\partial_i \gamma^e_b + \beta^c_i\partial_c\gamma^e_b 
- \gamma^c_b\partial_c\beta^e_i\right] 
\label{four} \,\,\,\,\,,\\
\xi^a_{ij} &=& \left(\gamma^{-1}\right)^a_e 
\left[\partial_i\omega^e_j -\partial_j\beta^e_i +\beta^b_i\partial_b\omega^e_j
-\omega^b_j\partial_b\beta^e_i\right] 
\label{five} \,\,\,\,\,,\\
\rho^a_i &=&
\left(\gamma^{-1}\right)^a_e 
\left[\omega^e_i - \beta^e_i\right]\,\,\,,
\label{six}
\end{eqnarray}
where $\partial_a$ and $\partial_i$ are, respectively, the derivatives with respect
to $y^a$ and $x^i$.
\par
Just as in the case of ordinary duality, we require that no explicit dependence
on the coordinates $y^a$ is left in the expression between brackets of the 
Bianchi identity (\ref{newbianchi}). 
That is, we demand that $\theta^a_b=\theta^a_b\left(x\right)$, $\varphi^a_{bc}=
\varphi^a_{bc}\left(x\right)$, $\mu^a_{ib}=\mu^a_{ib}\left(x\right)$,
$\nu^a_{ib}=\nu^a_{ib}\left(x\right)$, $\xi^a_{ij}=\xi^a_{ij}\left(x\right)$ and 
$\rho^a_i=\rho^a_i\left(x\right)$. 
Hence our task is to find the tensors $\alpha^a_b$, $\gamma^a_b$, $\omega^a_i$
and $\beta^a_i$ such that this property holds.
\par
We choose to express $\alpha^a_b\left(x,y\right)$ and $\omega^a_a\left(x,y\right)$
in terms of $\gamma^a_b\left(x,y\right)$ and $\beta^a_i\left(x,y\right)$. 
The first equation (\ref{one}) and the 
sixth equation (\ref{six}) of this set are algebraic and allow the determination 
of $\alpha^a_b$ and $\omega^a_i$. These are given by
\begin{eqnarray}
\alpha^a_b\left(x,y\right) &=& \gamma^a_c\theta^c_b 
\label{alpha} \,\,\,\,\,\,,\\
\omega^a_i\left(x,y\right)  &=& \beta^a_i
+ \gamma^a_c\rho^c_i\,\,\,.
\label{omega} 
\end{eqnarray}
Replacing $\alpha^a_b$
by its expression in equation (\ref{two}) leads to
\begin{equation}
\gamma^d_a\partial_d\gamma^c_b - \gamma^d_b\partial_d\gamma^c_a
=t^d_{ab}\gamma^c_d\,\,\,,
\label{liealgebra}
\end{equation}
where the new tensor $t^a_{bc}\left(x\right)$ is defined as
\begin{equation}
\theta^d_a t^c_{db}=\varphi^c_{ab}\,\,\,.
\end{equation}
Using (\ref{alpha}), (\ref{omega}), (\ref{four}) and (\ref{liealgebra}) in equation
(\ref{three}) we obtain
\begin{equation}
\mu^a_{ib} = -\partial_i\theta^a_b -\nu^a_{ic}\theta^c_b - t^a_{cd}\rho^c_i\theta^d_b
\,\,\,.
\label{mu}
\end{equation}
This last equation gives simply $\mu^a_{ib}$ in terms of the other tensors.  
Finally, using (\ref{omega}) and (\ref{four}) in equation
(\ref{five}) yields two relations: The first is
\begin{equation}
\partial_i\beta^a_j - \partial_j\beta^a_i + \beta^b_i\partial_b\beta^a_j
- \beta^b_j\partial_b\beta^a_i =
{1\over 2}\gamma^a_b\left(\xi^b_{ij} - \xi^b_{ji} - \partial_i\rho^b_j 
+\partial_j\rho^b_i - \nu^b_{ic}\rho^c_j + \nu^b_{jc}\rho^c_i\right)\,\,\,.
\label{curvature}
\end{equation}
The second relation is 
\begin{equation}
{1\over 2}\gamma^a_b\left(\xi^b_{ij} + \xi^b_{ji} - \partial_i\rho^b_j  
-\partial_j\rho^b_i - \nu^b_{ic}\rho^c_j - \nu^b_{jc}\rho^c_i\right)=0
\,\,\,.
\end{equation}
This last equation determines the symmetric part of $\xi^a_{ij}$ in terms 
of the other tensors. 
\par
The constraints on the two tensors $\gamma^a_b$ and $\beta^a_i$ are therefore
equations (\ref{liealgebra}), (\ref{four}) and (\ref{curvature}). These equations 
have a better interpretation in terms of the following differential operators
\begin{eqnarray}
T_a &=& \gamma^b_a\left(x,y\right)\partial_b\,\,\,,\nonumber\\
J_i &=& \beta^b_i\left(x,y\right)\partial_b\,\,\,.
\end{eqnarray}
Indeed, equations (\ref{liealgebra}), (\ref{four}) and (\ref{curvature}) become
respectively
\begin{eqnarray}
&& \left[T_a\,,\,T_b\right] = t^c_{ab} T_c \,\,\,,
\label{firstcommutator}\\
&& \left[\nabla_i\,,\,T_a\right] =\nu^b_{ia} T_b  \,\,\,,
\label{secondcommutator}\\
&& \left[\nabla_i\,,\,\nabla_j\right] =\Lambda_{ij}^a T_a\,\,\,,
\label{thirdcommutator}
\end{eqnarray}
where we have introduced the new notation
\begin{eqnarray}
\nabla_i &=& \partial_i + J_i \,\,\,,\nonumber\\
\Lambda_{ij}^a &=& {1\over 2}\left(\xi^a_{ij} - \xi^a_{ji} - \partial_i\rho^a_j 
+\partial_j\rho^a_i - \nu^a_{ic}\rho^c_j + \nu^a_{jc}\rho^c_i\right)\,\,\,.
\label{Lambda}
\end{eqnarray}
Of course, the above commutation relations have to satisfy certain consistency
relations. These are given by
\begin{eqnarray}
&& t^d_{ab}t^e_{dc} + t^d_{ca}t^e_{db}+ t^d_{bc}t^e_{da}=0\,\,\,,
\label{firstjacobi}\\
&& t^d_{ab}\nu^c_{id} - t^c_{ad}\nu^d_{ib}+ t^c_{bd}\nu^d_{ia}=-\partial_i t^c_{ab}\,\,\,,
\label{secondjacobi}\\
&& \partial_i\nu^a_{jb} - \partial_j\nu^a_{ib}
+ \nu^a_{ic}\nu^c_{jb} - \nu^a_{jc}\nu^c_{ib} = -t^a_{bd}\Lambda^d_{ij}\,\,\,,
\label{thirdjacobi}\\
&& \left(\partial_i \Lambda^a_{jk} + \nu^a_{ib}\Lambda^b_{jk}\right)
+  \left(\partial_k \Lambda^a_{ij} + \nu^a_{kb}\Lambda^b_{ij}\right)
+ \left(\partial_j \Lambda^a_{ki} + \nu^a_{jb}\Lambda^b_{ki}\right)=0\,\,\,.
\label{fourthjacobi}
\end{eqnarray}
These Jacobi identities do not require any further integrability conditions.
Furthermore, the last Jacobi identity is automatically satisfied once 
the first three are obeyed. It is worth mentioning that if the Jacobi identities
are fulfilled then the covariant derivative 
\bea
\left({\cal{D}}_i\right)^b_a &=& \delta^b_a\dd_i -\varepsilon^c_i t^b_{ca}
+\nu^b_{ia}\,\,\,,\nonumber\\
\varepsilon^a_i\left(x,y\right) &\equiv& \left(\gamma^{-1}\right)^a_b\beta^b_i
\eea
has vanishing curvature. Namely,
\be
\left({\cal{D}}_i\right)^a_b \left({\cal{D}}_j\right)^b_c - 
\left({\cal{D}}_j\right)^a_b \left({\cal{D}}_i\right)^b_c =0\,\,\,.
\label{DD=0}
\ee
This observation together with the above commutation relations give a geometrical 
interpretation to our setting.
\par
To summarise the problem so far, we are looking for two tensors $\gamma^a_b\left(x,y\right)$ 
and $\beta^a_i\left(x,y\right)$ from which we can construct the three tensors
$t^a_{bc}\left(x\right)$, $\Lambda^a_{ij}\left(x\right)$ and 
$\nu^a_{ib}\left(x\right)$. These last tensors are subject to the 
Jacobi identities (\ref{firstjacobi})-(\ref{fourthjacobi}).   

\section{Solutions}
By examining the first commutation relation (\ref{firstcommutator}) and the first Jacobi 
identity (\ref{firstjacobi}), one realises
that the tensor $t^a_{bc}$ must be related by a similarity transformation
to the structure constants of a Lie algebra. Namely, one must have
\be
t^a_{bc}\left(x\right)=s_b^d\left(x\right) s_c^e\left(x\right) S^a_g\left(x\right)f^g_{de}
\,\,\,,
\ee
where $S^a_b$ is the inverse of $s^a_b$ and $f^a_{bc}$ are the structure
constants of a Lie algebra $\cal{G}$, whose generators we denote $\lambda_a$,
such that
\be  
\left[\lambda_a\,,\,\lambda_b\right]=f^c_{ab}\lambda_c\,\,\,.
\ee
We deduce, from (\ref{liealgebra}), that $\gamma^a_b$ must be given by
\be
\gamma^a_b\left(x,y\right)=s_b^c\left(x\right)E^a_c\left(y\right)\,\,\,.
\ee
Here $E^a_b\left(y\right)$ are the inverses of the vielbeins $e^a_b\left(y\right)$ defined by
\be
e^a_b\lambda_a=g^{-1}\partial_b g\,\,\,,
\ee
where $g\left(y\right)$ is a Lie group element corresponding to the Lie algebra $\cal{G}$.
\par
Next, the expression for $\nu^a_{ib}$ in (\ref{four}) can be cast into
\be
\left(\nu^e_{ib} +s^d_b\partial_i S^e_d\right)S^b_a s^g_e=
\left(s^d_c\varepsilon^c_i\right)f^g_{da} -E^d_a\partial_d\left(s^g_c\varepsilon^c_i\right)
\,\,\,.
\ee
The right-hand-side of this equation must be independent of the coordinate $y^a$.
A possible solution to this problem is found by taking
\be
\varepsilon^c_i\left(x,y\right)=S^c_b\left(x\right)\left[v^b_a\left(y\right) 
\kappa^a_i\left(x\right)
+{\cal{C}}^b_i\left(x\right)\right]\,\,\,,
\ee
where $v^a_b\left(y\right)$ is defined as
\be
v^a_b\lambda_a=g^{-1}\lambda_b g
\ee
and satisfies the following crucial properties
\be
E^d_c\partial_dv^e_a -v^b_af^e_{bc}=0\,\,\,\,\,\,\,\,\,\,,\,\,\,\,\,\,\,
v^b_av^d_cf^e_{bd}-v^e_df^d_{ac}=0\,\,\,.
\ee
This solution leads to the following final expression for the tensor $\nu^a_{ib}\left(x\right)$
\be
\nu^e_{ib}\left(x\right)=-s^a_b\left(\partial_iS^e_a - f^g_{da}{\cal{C}}^d_i S^e_g\right)
\,\,\,.
\ee
What remains to be determined now is the tensor $\xi^c_{ij}\left(x\right)$.
Equation (\ref{Lambda}) leads to 
\be
\Lambda^c_{ij}\left(x\right) = \partial_i\varepsilon^c_j - \partial_j\varepsilon^c_i
-t^ c_{eb}\varepsilon^e_i \varepsilon^b_j +\nu^c_{ie}\varepsilon^e_j - \nu^c_{je}\varepsilon^e_i
\,\,\,.
\ee
Replacing for $\varepsilon^c_i$, $t^c_{ab}$ and $\nu^a_{ib}$ and using the 
properties of $v^a_b$ one gets
\be
\Lambda^c_{ij}\left(x\right) = v^b_a\left(y\right) S^c_b\left[
\dd_i\kappa^a_j - \dd_j\kappa^a_i - f^a_{de}\kappa^d_i \kappa^e_j\right]
+ S^c_a\left[\dd_i{\cal{C}}^a_j - \dd_j{\cal{C}}^a_i +
f^a_{de}{\cal{C}}^d_i {\cal{C}}^e_j\right]\,\,\,.
\ee
Hence, the right-hand-side is independent of $y^a$ only and only if
\be
\dd_i\kappa^a_j - \dd_j\kappa^a_i - f^a_{de}\kappa^d_i \kappa^e_j =0\,\,\,.
\label{zerocurvature}
\ee
An interesting solution to this condition is given by 
\be
\kappa^a_i\left(x\right) \lambda_a = \partial_i h h^{-1}\,\,\,,
\ee 
where $h(x)$ is an element of a Lie group built on the Lie algebra
$\cal{G}$ but with parameters $x^i$. Of course, this solution requires that
the range of the indices $\left\{a,b,c,\dots\right\}$ is the same as that of a subset of the indices
$\left\{i,j,k,\dots\right\}$. If the Lie algebra ${\cal{G}}$ is Abelian then $v^a_b=\delta^a_b$ and
the zero curvature condition (\ref{zerocurvature}) is not necessary.  
\par
Finally, the expression of the tensor $\xi^c_{ij}\left(x\right)$
is given by
\be
\xi^c_{ij}\left(x\right)= S^c_a\left[\dd_i{\cal{C}}^a_j - \dd_j{\cal{C}}^a_i +
f^a_{de}{\cal{C}}^d_i {\cal{C}}^e_j\right]
+
\dd_i\rho^c_j -s^e_d\left(\dd_i S^c_e -f^g_{be}{\cal{C}}^b_i S^c_g\right)\rho^d_j
\,\,\,.
\ee  
It is easy to check that all the Jacobi identies are satisfied by the solution
presented here.

\section{The dual models}

The original two-dimensional non-linear sigma model is still given by an action 
of the form (\ref{original})
but with $A^a$ and $\bar A^a$ as expressed by
\bea
A^a &=& S^c_d\left(\theta^{-1}\right)^a_c\left[e^d_b\dd y^b -\left(v^d_e\kappa^e_i 
+{\cal{C}}^d_i\right)\dd x^i\right]\nonumber\\
\bar A^a &=& S^a_d\left[e^d_b \bar\dd y^b -\left(v^d_e\kappa^e_i 
+{\cal{C}}^d_i +s^d_e\rho^e_i\right)\bar\dd x^i\right]\,\,\,.
\eea
We now make use of the Bianchi identity corresponding to this gauge fields 
to write the first order action
\begin{eqnarray}
I_1\left[x,A,\bar A, \chi\right] & = & 
\int {\rm{d}}z{\rm{d}}\bar{z}\left\{
Q_{ij}\left(x\right)\partial x^i \bar\partial x^j
+G_{ia}\left(x\right)\partial x^i \bar{A}^a  
+\bar G_{ia}\left(x\right)\bar\partial x^i A^a \right. \nonumber \\
&& + 
P_{ab}\left(x\right) A^a \bar{A}^b 
+ \chi_b\left[\partial\bar A^b - \theta^b_c\left(x\right)\bar\partial A^c
+ \varphi^b_{cd}\left(x\right)A^c\bar A^d 
+ \mu^b_{ic}\left(x\right)\bar\partial x^i A^c \right.\nonumber\\ 
&&
+\nu^b_{ic}\left(x\right)\partial x^i \bar A^c
+ \xi^b_{ij}\left(x\right)\partial x^i \bar\partial x^j
+\rho^b_i\left(x\right)\partial \bar \partial x^i \left.\right]
\left.\right\}\,\,\,. 
\end{eqnarray}
The original action is obtained by substituting for $A^a$ and $\bar A^a$.
On the other hand, the dual theory is obtained by keeping the Lagrange multiplier
$\chi_a$ and integrating out the gauge fields. It is, however, possible to give 
a simple form to the two dual models.
\par
In order to compare our construction with ordinary non-Abelian duality,
we introduce the new one-form defined by\footnote{I am grateful to Ian jack for 
suggesting this to me.}
\bea
B^a = s^a_b\theta^b_c A^c +{\cal{C}}^a_i\dd x^i
&=&
e^a_b\left(y\right)\dd y^b - v^a_b\left(y\right)\kappa^b_i\left(x\right)\dd x^i
\nonumber\\
\bar B^a = s^a_b\bar A^b +\left({\cal{C}}^a_i +s^a_b\rho^b_i\right)\bar\dd x^i
&=&
e^a_b\left(y\right)\bar\dd y^b - v^a_b\left(y\right)\kappa^b_i\left(x\right)\bar\dd x^i
\,\,\,.
\label{BB}
\eea
It is then easy express the Bianchi identity associated to $A^a$ and $\bar A^a$ in 
terms of the new gauge fields $B^a$ and $\bar B^a$. This takes the familiar form 
\begin{equation}
E^a_b \left(\partial \bar B^{b} - \bar \partial B^{b}
+f_{cd}^{b}B^{c}\bar B^{d}\right)=0\,\,\,.
\end{equation}
Indeed, using group elements, $B^a$ and $\bar B^a$ are such that
\be
B^a \lambda_a = l^{-1}\dd l\,\,\,\,,\,\,\,\, 
\bar B^a \lambda_a = l^{-1}\bar \dd l\,\,\,\,,
\ee
where $l\left(x,y\right)=h^{-1}\left(x\right)g\left(y\right)$. This remark explains the 
zero curvature condition encountred in (\ref{DD=0}).
\par
In terms of the gauge fields $B^a$ and $\bar B^a$, the above first order action
takes the form
\begin{eqnarray}
I_1\left[x,B,\bar B, \chi\right] & = & 
\int {\rm{d}}z{\rm{d}}\bar{z}\left\{
T_{ij}\left(x\right)\partial x^i \bar\partial x^j
+H_{ia}\left(x\right)\partial x^i \bar{B}^a  
+\bar H_{ia}\left(x\right)\bar\partial x^i B^a \right. \nonumber \\
&& + 
\left. F_{ab}\left(x\right) B^a \bar{B}^b 
+ \chi_b\left(\partial \bar B^{b} - \bar \partial B^{b}
+f_{cd}^{b}B^{c}\bar B^{d}\right)
\right\}\,\,\,.
\label{newfirstorder}
\end{eqnarray}
Here, the new backgrounds $T_{ij}$, $H_{ia}$, $\bar H_{ia}$ and $F_{ab}$ are functions 
of the old tensors  $Q_{ij}$, $G_{ia}$, $\bar G_{ia}$ and $P_{ab}$.
The original sigma model is then obtained by solving the constraints imposed by the
Lagrange multiplier $\chi_a$. This amounts to replacing, in the first order action, 
$B^a$ and $\bar B^a$ by their experssions in (\ref{BB}). We obtain the original theory
\begin{eqnarray}
I\left[x,y\right] &=& 
\int {\rm{d}}z{\rm{d}}\bar{z}\left\{
\left(T_{ij}  + F_{ab}v^a_cv^b_d\kappa^c_i\kappa^d_j
-H_{ia}v^a_b\kappa^b_j - \bar H_{ja}v^a_b\kappa^b_i\right)\partial x^i \bar\partial x^j
\right.
\nonumber\\
&&+\left(H_{ia}-F_{ba}v^b_c\kappa^c_i\right)e^a_d\partial x^i \bar \dd y^d  
+\left(\bar H_{ia} - F_{ab}v^b_c\kappa^c_i\right)e^a_d\bar\partial x^i \dd y^d
\nonumber\\ 
&& + F_{ab}e^a_ce^b_d \dd y^c\bar\dd y^d \left.\right\}\,\,\,.
\label{neworiginal}
\end{eqnarray}
Notice that all the dependance on the $y^a$ coordinates is in $e^a_b(y)$ and $v^a_b(y)$.
It is clear that if $\kappa^a_i$ is set to zero then one recovers the setting for ordinary 
non-Abelian duality ($f^a_{bc}\neq 0$) or Abelian duality ($f^a_{bc}=0$). 
\par
The dual non-linear sigma model is found by eliminating, through their equations of motion,
the gauge fields $B^a$ and $\bar B^a$ from the first order action. This leads to a dual theory 
described by
\begin{eqnarray}
\widetilde{I}\left[x,\chi\right] &=&  \int {\rm{d}}z{\rm{d}}\bar{z}\left\{
\left[T_{ij} - \left(M^{-1}\right)^{ab} H_{ia} \bar H_{jb}\right]
\partial x^i \bar\partial x^j
- \left(M^{-1}\right)^{ab} H_{ia} \dd x^i \bar\dd\chi_b
\right.\nonumber\\
&&+\left(M^{-1}\right)^{ba} \bar{H}_{ia} \bar\dd x^i \dd\chi_b
+\left(M^{-1}\right)^{ab}\partial\chi_a\bar\partial\chi_b
\left.\right\}\,\,\,,
\label{newdual} 
\end{eqnarray} 
where $M_{ab}=F_{ab}+\chi_cf^c_{ab}$.

\section{Discussion}

The construction associated to Abelian and non-Abelian T-duality has been 
generalised to find new T-duality transformations. The analyses has been carried out 
at the level of the classical non-linear sigma model action. 
It relies on the integrability condition (Bianchi identity) associated to some one-form
${\cal{A}}^a=A^a{\rm d}z+\bar A^a{\rm d}\bar z$. The duality procedure can be carried out
if the dependance of the original action on some fields $y^a$ appears through 
this one-form only. One can then trade the coordinates $y^a$ for the one-form ${\cal{A}}^a$
at the price of introducing a Lagrange multiplier. The equations of motion for this Lagrange multiplier
yield the original theory. The gauge fields $A^a$ and $\bar A^a$ appear quadratically in the action and without 
derivatives. Their equations of motion lead to a dual sigma model in which the Lagrange multiplier
replaces the coordinates $y^a$. 
\par
A natural question in this context is whether this new T-duality remains a symmetry at the quantum 
level (see \cite{our1} and references within for a quantum treatment of Abelian and non-Abelian 
T-dualities).
In other words, we would like to know if the string effective action corresponding to the
backgrounds of the model (\ref{neworiginal}) is equivalent to that coming from the
backgrounds  of the action (\ref{newdual}). This study requires a knowledge of the transformation
properties of the dilaton field. If the two-dimensional theory (\ref{neworiginal}) possesses a dilaton 
background in the form 
\be
-{1\over 4}\int {\rm d}^2\sigma\sqrt\eta R^{(2)} \Phi\left(x\right)\,\,\,,
\ee
where $\eta$ is the metric on the world-sheet and $R^{(2)}$ is its Ricci scalar curvature.
In analogy with Abelian and non-Abelian T-duality \cite{dilaton}, we conjecture that the dilaton in the dual
theory is given by
\be
\widetilde{\Phi} = \Phi -{1\over 2} \ln \det \left(M\right)\,\,\,.
\ee 
The study of the string effective action under these new T-duality transformation
is worth exploring. 
\par
Finally, we should mention that a duality transformation based on 
the gauging of the most general sigma model having a chiral symmetry
($H_{\rm L}\times H_{\rm R}$) was found in ref.\cite{obers1}
and some of its applications were explored in \cite{obers2}.
There, different kinds of gauging (a generalisation of Abelian axial and vector gauging)
were used to construct the duality transformations. This duality symmetry reduces to ordinary Abelian 
duality if the chiral group is Abelian.
The relation between this ``axial-vector'' duality symmetry and non-Abelian duality (or our new duality)
remains undetermined.

\vskip 2 cm
\noindent
{\bf Aknowledgments:}
I would like to thank Janos Balog,  Peter Forg\'acs, Ian Jack 
and Max Niedermaier for useful discussions.

\end{document}